\documentclass[lettersize,journal]{IEEEtran}

\usepackage{epsfig,scalefnt,multirow}
\usepackage{url}
\usepackage{ifthen}
\usepackage{mathtools}
\usepackage{cite}
\usepackage{graphicx}
\usepackage{amssymb}
\usepackage{caption}
\usepackage{booktabs}
\usepackage{amsmath}
\usepackage{epstopdf}
\usepackage{algorithm}
\usepackage{algpseudocode}
\usepackage{cases}
\usepackage{xcolor}
\usepackage{multicol}
\usepackage{stfloats}
\usepackage{blindtext}

\def\b0{{\pmb{0}}} 

\begin{document}

\title{Machine Learning for Future Wireless Communications: Channel Prediction Perspectives}

\author{Hwanjin Kim,~\IEEEmembership{Member,~IEEE,} Junil Choi,~\IEEEmembership{Senior Member,~IEEE,} David J. Love,~\IEEEmembership{Fellow,~IEEE}
\thanks{H. Kim is with the School of Electronics Engineering, Kyungpook National University, Daegu 41566, South Korea (e-mail: hwanjin@knu.ac.kr).}
\thanks{J. Choi is with the School of Electrical Engineering, Korea Advanced Institute of Science and Technology, Daejeon 34141, South Korea (e-mail: junil@kaist.ac.kr).}
\thanks{D. J. Love is with the Elmore Family School of Electrical and Computer Engineering, Purdue University, West Lafayette, IN 47907 USA (e-mail: djlove@purdue.edu).}}

\maketitle

\begin{abstract}
Precise channel state knowledge is crucial in future wireless communication systems, which drives the need for accurate channel prediction without additional pilot overhead. While machine-learning (ML) methods for channel prediction show potential, existing approaches have limitations in their capability to adapt to environmental changes due to their extensive training requirements. In this paper, we introduce the channel prediction approaches in terms of the temporal channel prediction and the environmental adaptation. Then, we elaborate on the use of the advanced ML-based channel prediction to resolve the issues in traditional ML methods. The numerical results show that the advanced ML-based channel prediction has comparable accuracy with much less training overhead compared to conventional prediction methods. Also, we examine the training process, dataset characteristics, and the impact of source tasks and pre-trained models on channel prediction approaches. Finally, we discuss open challenges and possible future research directions of ML-based channel prediction.
\end{abstract}


\section{Introduction}
Machine learning (ML) algorithms are thought to hold great potential in nearly all future wireless communication systems due to their adaptability and high performance. In wireless systems, the channel between the base station (BS) and the user equipment (UE) is affected by various factors, e.g., multipath fading, mobility, interference, and noise \cite{Truong13}. To obtain accurate channel state information (CSI), the BS needs to use pilots, which incurs the overhead. To mitigate the overhead, the BS can predict future channel realizations by leveraging historical channel measurements \cite{Choi14}.

Channel prediction can tackle many of the challenges found in emerging wireless communications. Accurate CSI acquisition in reconfigurable intelligent surface (RIS) systems is challenging due to the large number of elements in RIS, leading to high-dimensional channel prediction complexity and increased computational overhead. Similarly, cell-free massive MIMO systems face a channel aging issue due to the high mobility of UEs, which further degrades performance depending on the access point (AP) deployment environment and the synchronization among distributed APs. Satellite communications, which are of much interest for 6G, offer unique CSI issues arising from the long transmission delay, time-varying channel due to extremely high mobility, and limited onboard computational resources, making real-time adaptation more challenging.
	
In wireless communications, there are two main categories for channel prediction: classical model-based approaches and data-driven ML-based approaches. Model-based channel prediction relies on mathematical models that describe the behavior of the wireless channel. These models are typically based on physical or statistical assumptions about propagation environments. The statistical models for channel prediction usually rely on an auto-regressive (AR) model, which captures the statistical properties and temporal dependencies of the channel, followed by Kalman filtering, which handles the noise and uncertainties of the measurements \cite{Shikur15}. Deterministic channel models, such as ray-tracing, provide highly accurate, site-specific predictions by considering exact geometries and material properties of the environment. Although effective, model-based approaches often require prior knowledge of the environment and assumptions about the channel statistics, which may not capture the complex and non-linear properties of time-varying channels found in real-world deployments.

Over the last few years, data-driven channel prediction techniques using ML methods have been proposed for future wireless communications in \cite{Wu21}. Channel prediction using a multi-layer perceptron (MLP) was developed with linear minimum mean square error (LMMSE) pre-processing in \cite{Kim21}, and a channel predictor using a convolutional neural network (CNN)-AR and the auto-correlation pattern was discussed in \cite{Yuan20}. Also, a recurrent neural network (RNN) with long short-term memory (LSTM)-based channel prediction was developed in \cite{Jiang20}. In addition, a parallel vector mapping model-based channel prediction using an attention mechanism was proposed in \cite{Jiang22}. With the attention mechanism, the relationships through historical channels can be captured to improve the prediction performance. Data-driven approaches can adapt to different environments and capture non-linear relationships from a sufficient amount of training data, but their effectiveness heavily depends on the accessibility and quality of the training data. These approaches also encounter difficulties when dealing with unseen channel conditions, e.g., a UE newly connected to the BS. To deal with these unseen channel conditions, more advanced ML techniques may be required.

\begin{figure*}[!t]
	\centering
	\includegraphics[width=5.5 in]{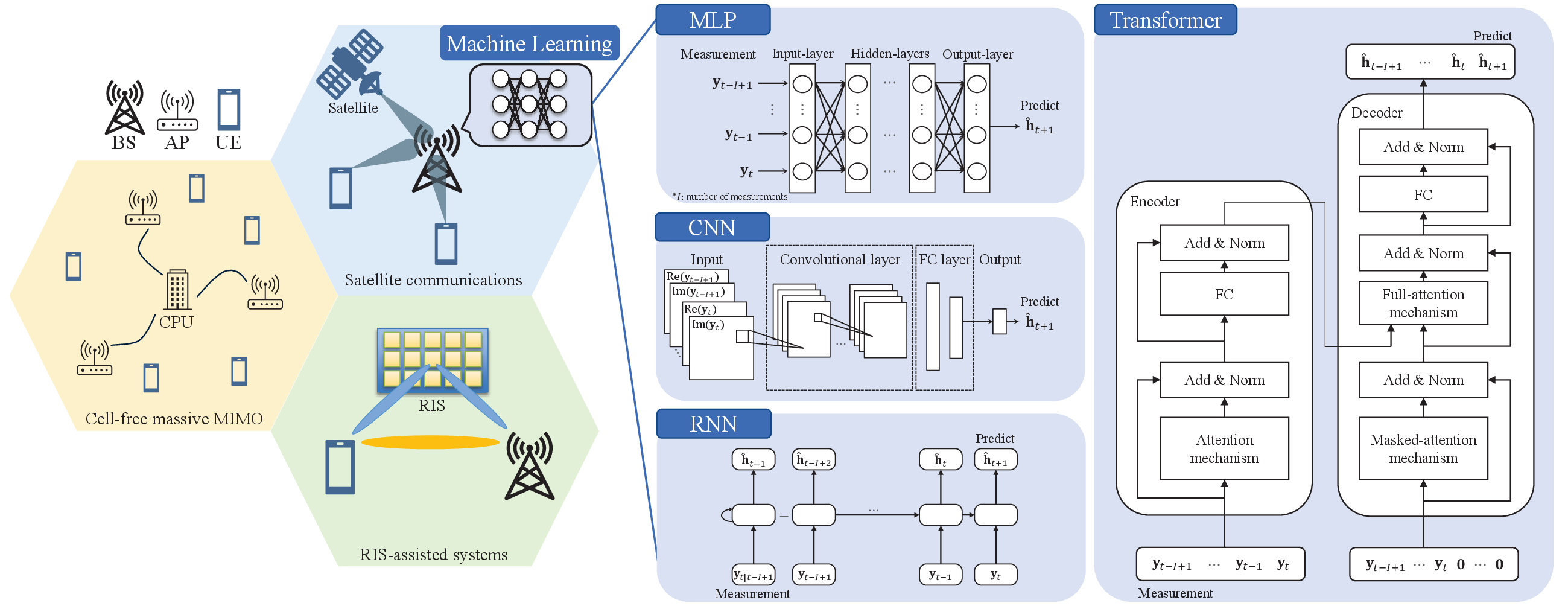}
	\caption{Channel prediction for future wireless communications with ML-based approaches: MLP, CNN, RNN models, or attention mechanism (transformer in [8]).}
	\label{fig_1}
\end{figure*}

Advanced ML-based techniques such as transfer learning-based and meta-learning-based schemes have been developed in \cite{Yuan21, Kim23}. Both transfer learning and meta-learning are designed to improve the generalization of models to unseen tasks by leveraging the information learned from previous tasks.
In the transfer learning and meta-learning jargons, previous tasks and unseen tasks are often referred to as source tasks and target tasks, respectively.
Instead of training a model from scratch, both learning-based schemes leverage the knowledge learned from the source dataset, i.e., the channel used to train a pre-existing model, to predict the target dataset, i.e., the channel for a specific task not used for training the pre-existing model, which can significantly reduce the training overhead for a target task.
Also, data augmentation, accomplished by applying transformations to the original data, can be used to artificially enhance the dataset \cite{Raviv23}. Data augmentation for channel prediction mainly focuses on diversity of the dataset of CSI to make the prediction models more robust and capable of generalizing to different environments. Generative channel estimators, which can leverage underlying algorithms such as generative adversarial networks (GANs) or diffusion models, hold much potential to generate representative channel estimates. This strategy can play a significant role in enhancing the performance of predictive models, especially when training data are limited.

In this paper, we clearly distinguish between temporal channel prediction and environmental adaptation. Temporal channel prediction focuses on predicting future channel states based on past observations under relatively stable environmental conditions, which leverages the temporal correlation of the channel. In contrast, environmental adaptation aims to adjust the prediction model to accommodate changes in surrounding environments, allowing the model to maintain robust performance even in dynamic and previously unseen scenarios. We first overview the basic ideas behind temporal prediction of MIMO channels and motivate environmental adaptation. We also introduce the classical model-based and ML-based channel prediction approaches according to the channel prediction types. For environmental adaptation, we reveal that the model-based channel prediction approaches have a fundamental limitation to achieving accurate performance and elaborate on how to handle these issues with advanced ML-based approaches, e.g., transfer learning, meta-learning, and data augmentation. Finally, we discuss possible challenges and future research directions for each challenge.

\section{System Model and Channel Prediction Architecture}
As shown in Fig. \ref{fig_1}, ML-based channel prediction techniques can be applied to future wireless communications, e.g., RIS-assisted systems, cell-free massive MIMO, and satellite communications. In this paper, we first focus on basic temporal channel prediction. To predict the temporal channel, a classical model-based channel predictor, e.g., Kalman filtering based on the AR model or an ML-based channel prediction exploiting MLP, CNN, RNN models, or attention mechanism, can be used.

\begin{figure*}[!t]
	\centering
	\includegraphics[width=4.5 in]{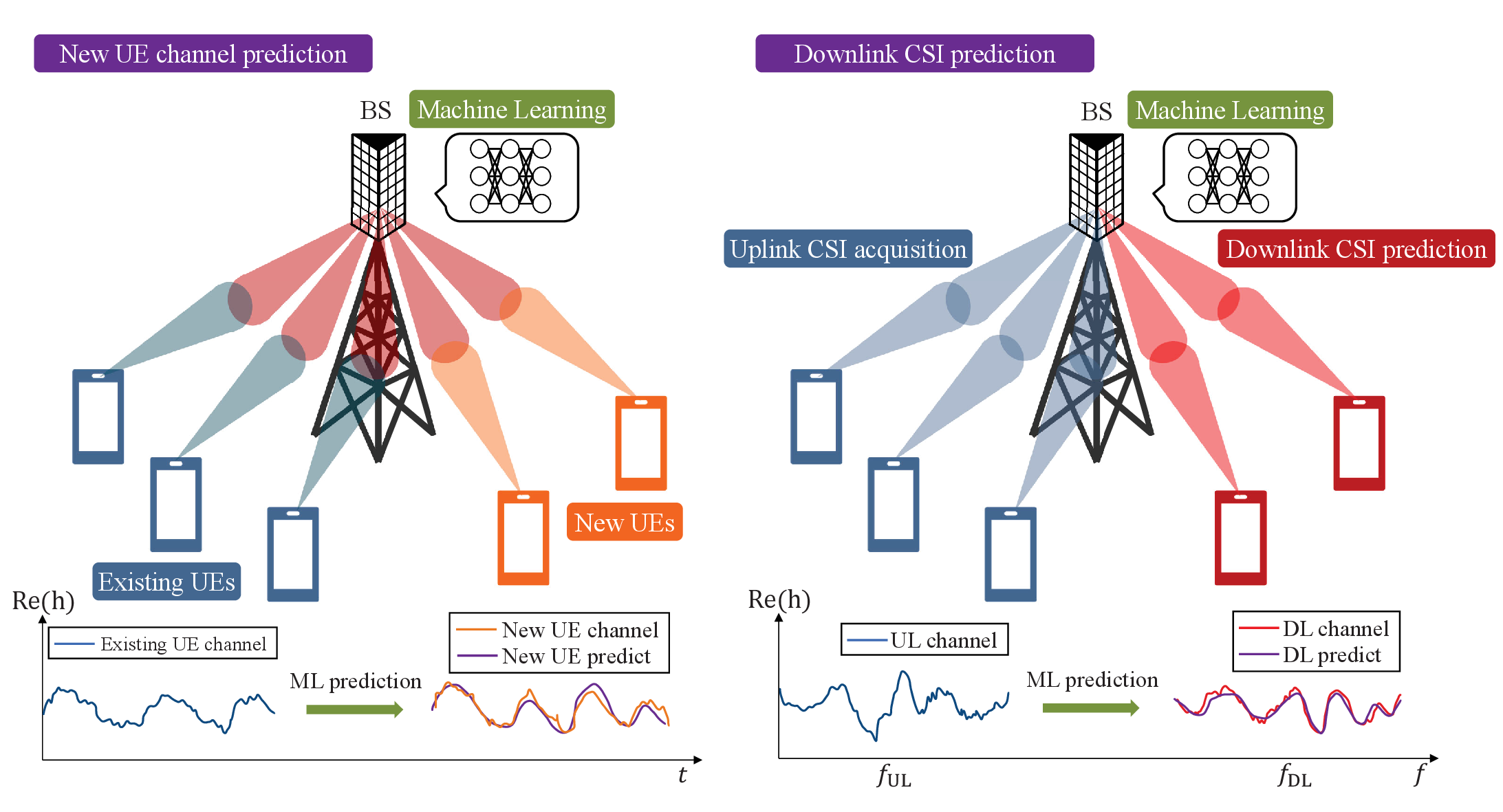}
	\caption{Environmental adaptation: New UE channel prediction and downlink CSI prediction.}
	\label{fig_2}
\end{figure*}

However, previous model-based and simple ML-based approaches cannot be directly applied to more sophisticated scenarios, such as environmental adaptation. 
In contrast to non-temporal and environmental agnostic models, which reflect relatively static characteristics and typically require estimation, our work on environmental adaptation focuses on prediction techniques for channels influenced by both dynamic temporal variations and environmental changes. Contrary to temporal channel prediction alone, environmental adaptation must take into account the domain mismatch between the source dataset and target dataset. Therefore, it can involve various scenarios, e.g., new UE channel prediction based on existing UE channels \cite{Kim23} or downlink CSI prediction based on uplink CSI \cite{Yang20} described in Fig. \ref{fig_2}. 
For channel prediction in the environmental adaptation, transfer learning, meta-learning, and data augmentation can be employed.

Advantages and limitations of channel prediction approaches depending on the two types of channel predictions, i.e., temporal channel prediction and environmental adaptation, are summarized in Table I. We also provide the practical guidance and corresponding model selection in Table II. Then, the detailed explanation follows in Sections \ref{Section3} and \ref{Section4}. 

\begin{table*}[t]
	\centering
	\caption{Features of channel prediction approaches depending on prediction types.}
	\includegraphics[width=5.5 in]{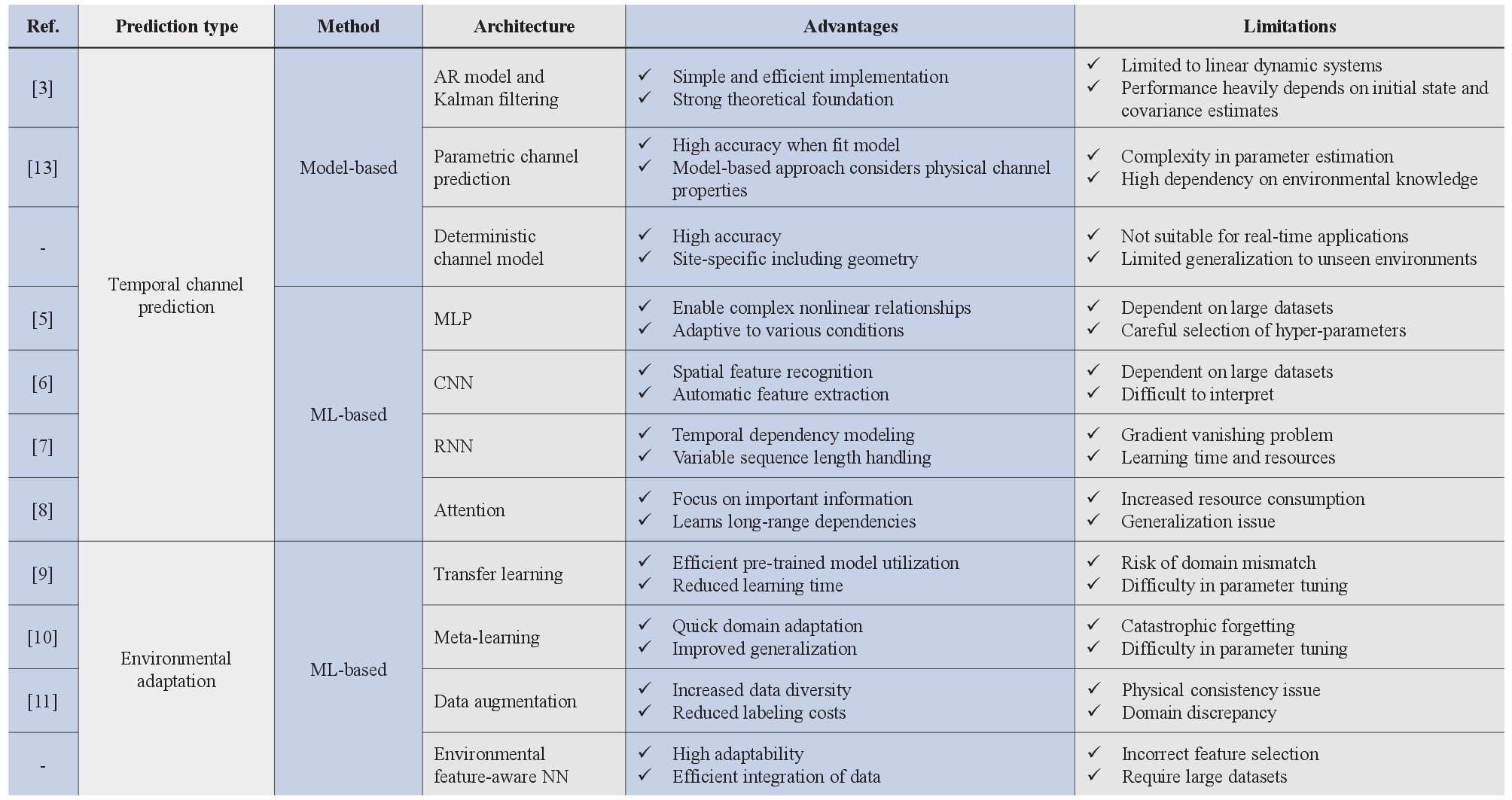}
	\label{table1}
\end{table*}

\begin{table*}[t]
	\centering
	\caption{Practical guidance and model selection.}
	\includegraphics[width=5.5 in]{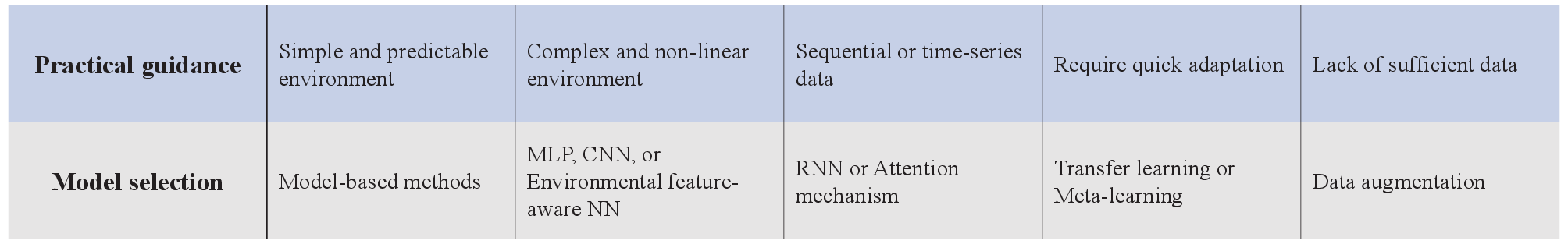}
	\label{table2}
\end{table*}

\section{Temporal Channel Prediction}\label{Section3}
In this section, we first elaborate on the model-based temporal channel prediction, e.g., AR model, Kalman filter-based prediction, parametric channel prediction, and deterministic channel model. Then, we discuss ML-based prediction, which includes MLP, CNN, RNN, and attention-based prediction. 

\subsection{AR model and Kalman filter-based channel prediction}
In an AR model, the sample at the current time can be written as a linear combination of the previous time samples and a stochastic innovation. Channel prediction using the AR model can be formulated by the AR-order, AR parameter matrix, and innovation process \cite{Shikur15}. To determine the appropriate AR-order, we can use the autocorrelation function (ACF) or Akaike information criterion (AIC). 
The AR model parameters can be estimated using the Yule-Walker equations, which minimize the difference between observed and predicted values. Once the AR model is fitted, the Kalman filter (KF) predicts CSI by combining weighted current and past CSI. Note that the AR model assumes linearity and may not capture non-linear properties of wireless channel. Therefore, the effectiveness may vary depending on wireless channel environments.

\subsection{Parametric channel prediction}
Parametric channel prediction models a channel as a superposition of complex sinusoidal functions, which have amplitudes, phases, and Doppler shifts based on a traditional stochastic channel model \cite{Adeogun13}. 
Each scattering source contributes to the channel response, requiring the estimation of its amplitude, Doppler shift, angle of arrival, and angle of departure.

The procedure for fading channel parameter estimation involves multiple stages. First, the covariance matrix is estimated, followed by determining the number of dominant scattering sources. Classical algorithms, e.g., multiple signal classification (MUSIC), are then used to estimate Doppler shifts and angles of arrival and departure. Also, the amplitude of each path can be calculated based on the estimated parameters and the number of dominant scattering sources. These values are plugged into a parametric channel model for prediction. However, this process is computationally intensive, and the parameters lose validity quickly in high-mobility environments, requiring frequent re-estimation, which can be impractical.

\subsection{Deterministic channel model}
Deterministic channel models, as a part of the model-based prediction approaches, offer a precise method for modeling wireless channels by following the laws of physics. Unlike stochastic or empirical models, deterministic models such as ray-tracing models provide site-specific predictions by considering the exact geometry and material properties of the environment. This approach is highly accurate and offers detailed insights into signal propagation, capturing complex interactions such as reflection, diffraction and scattering. However, deterministic channel models come with limitations. Their computational complexity is a major drawback, as they require substantial resources for simulation, making them unsuitable for real-time applications. Moreover, the accuracy of deterministic models heavily relies on the precision of input data, such as environmental dimensions and material properties.

\subsection{Neural network-based prediction}
While any neural network (NN) structure can be used for channel prediction, we elaborate on the representative NNs, e.g., MLP, CNN, RNN, and attention mechanism as in Fig. \ref{fig_1}. Channel prediction using MLP leverages the feed-forward architecture to model complex nonlinear relationships in channel data \cite{Kim21}. By feeding in current and past measurements, MLP predicts future channel states, with training typically minimizing the mean square error (MSE) between predicted and actual values. While MLP has the advantage of being able to capture nonlinear relation in the channel data, MLP often requires significant data and careful tuning of hyper-parameters to achieve optimal performance.

Channel prediction using CNNs effectively captures spatial and temporal dependencies in channel data \cite{Yuan20}. The CNN model architecture for channel prediction would consist of convolutional, pooling, and fully-connected (FC) layers. For a suitable format for CNN input, for example, we can represent the channel data as a 2D image with antenna domain as rows and sub-carriers as columns. Additionally, CNN models can be combined with AR models, where CNNs extract aging patterns, and AR predicts future CSI.

RNNs, particularly LSTMs, are effective for channel prediction due to their ability to capture temporal dependencies in sequential data \cite{Jiang20}. The LSTM models have specialized units that help in capturing long-term dependencies in the sequence. RNNs can suffer from the vanishing or exploding gradient problem, where the gradients used for weight updates become extremely small or large during the training. The LSTM units help mitigate this issue, but it can still impact the performance of the model. However, RNNs, including LSTMs, can be computationally intensive, especially for large-scale datasets and complex models, which presents challenges for real-time systems or resource-constrained devices.

Attention mechanisms, widely used in machine learning, can also be applied in wireless communications for channel prediction \cite{Jiang22}. The attention mechanism can be used to weigh the relevance of different historical data points during the prediction. When combined with RNNs or LSTMs, attention helps prioritize critical input sequences, capturing long-range dependencies and mitigating vanishing gradient issues. Self-attention or scaled dot-product attention can focus on any part of the input sequence, capturing long-range dependencies, which can mitigate the vanishing problems in RNNs. This leads to more accurate predictions, especially in complex environments, by focusing on the most relevant information. However, this attention model may struggle to generalize to new environments, such as different frequency bands of channels.

\section{Environmental Adaptation}\label{Section4}
In this section, we elaborate in detail on how the advanced ML architectures can resolve the environmental adaptation problems. We introduce transfer learning, meta-learning, data augmentation, and environmental feature-aware NN, then evaluate the performance of environmental adaptation approaches. 

\subsection{Transfer learning}
Transfer learning-based channel prediction starts with a model trained on a large dataset (typically not related directly to channel prediction). The model is then fine-tuned for the specific task of predicting the behavior of a wireless channel \cite{Yuan21}. Models like auto-encoders or LSTMs can serve as starting points. For adaptation, the final layers are replaced with task-specific ones, and the pre-trained layers are frozen during training. After training the new layers, the pre-trained layers can optionally be unfrozen for further fine-tuning. The main advantage of using transfer learning for channel prediction is that it can leverage the ability of the pre-trained model to extract useful features from the training data. This potentially leads to better performance even when the amount of channel prediction data is limited.

\subsection{Meta-learning}
Meta-learning focuses on learning to learn by training models across multiple source tasks to improve their ability to quickly adapt to new target tasks \cite{Kim23}. Model agnostic meta-learning (MAML), a popular meta-learning approach, uses a hierarchical structure with a meta-learner optimizing the learning process and a learner adapting to the target task. When faced with a target task, e.g., new UE channel prediction or downlink CSI prediction as in Fig. \ref{fig_2}, the learner uses the information from the meta-learner (with the initialization) to rapidly adapt to a new environment. This hierarchical structure in meta-learning is powerful since it divides into task-specific learning from the higher-level learning process, enabling models to quickly adapt to  target tasks with limited data. Therefore, meta-learning is useful in the scenarios where the channel prediction data may be scarce to obtain. Furthermore, the Bayesian online meta-learning framework can mitigate the catastrophic forgetting during the training phase.

\textit{Remark}: the main difference between transfer learning and meta-learning is in their adaptability. Transfer learning fine-tunes a model trained on a large source task for a specific target task, but it requires re-training for new tasks. Meta-learning, however, trains a model to quickly adapt to new tasks using minimal data. While transfer learning models need fine-tuning for each new task, meta-learning models can handle a range of tasks without extensive retraining.

\subsection{Data augmentation}
Data augmentation approaches, such as those that use a GAN or variational auto-encoder (VAE), are used in ML to artificially increase the size of dataset by introducing slight modifications to the existing data \cite{Raviv23}. Data augmentation can help improve the data diversity, making them less prone to over-fitting, and can be reduced the labeling costs in CSI. The generative models, e.g., GAN or VAE, are used to produce synthetic CSI that resemble the real data. 
The GAN-based method has the capacity for extension into a conditional GAN (cGAN) framework, which enables learning the conditional probability distribution of the data samples. Consequently, cGANs are capable of capturing the statistical distribution of samples and conforming to the constrained relationship between input and output data.
With a richer and more varied dataset, the generalization performs better to unseen scenarios. Also, the models become more robust to real-world inconsistencies by simulating various disturbances and noise in the data. Thus, the data augmentation can be a valuable tool in channel prediction to improve model robustness and performance, especially when faced with limited real-world CSI data.

\begin{figure}[!t]
	\centering
	\includegraphics[width=3.2 in]{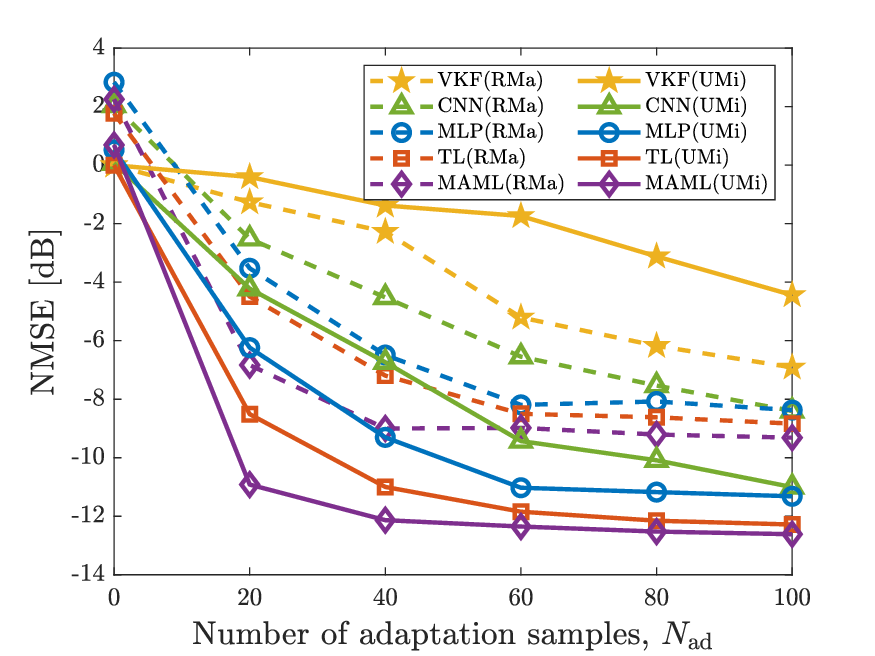}
	\caption{NMSE vs. number of adaptation samples in environmental adaptation.}
	\label{fig_3}
\end{figure}

\subsection{Environmental feature-aware NN}
Environmental feature-aware NNs are a class of models designed to capture complex environmental features such as building layouts, vegetation, or urban structures, which significantly impact wireless signal propagation. These models are highly adaptable to dynamic environments, making them effective for wireless channel prediction in scenarios where signal interference and reflections depend on the surroundings. By incorporating diverse data inputs like satellite images, LiDAR scans, and detailed environmental maps, these NNs can generate more accurate representations of real-world scenarios. This adaptability is especially crucial in urban and indoor environments, where obstacles like buildings and walls can substantially affect signal propagation. While environmental feature-aware NNs improve prediction by leveraging relevant features, challenges arise from feature selection and data requirements, as improper selection or excessive diversity can degrade performance. Also, training these models demands significant computational resources and large amounts of labeled data to ensure optimal performance.

\subsection{Numerical results for environmental adaptation}
In Fig. \ref{fig_3}, we evaluate the normalized mean square errors (NMSEs) of vector Kalman filter (VKF), CNN, MLP, transfer learning (TL), and MAML-based channel predictions according to number of adaptation samples in the 3GPP spatial channel model (SCM) for an urban micro (UMi) scenario and a rural macro (RMa) scenario.  We assume that the BS predicts the new UE channel based on the training samples from existing UEs (source tasks) and the adaptation samples from new UEs (target tasks). We use 1024 training samples per UE, then we test on 100 samples. Details of all simulation setups can be found in \cite{Kim23}. CNN and MLP re-train the whole network parameters, and TL only fine-tunes the last FC layer of the model in the adaptation phase. The NMSEs of all channel predictions decrease when the number of adaptation samples increases. Channel predictions using MLP, TL, and MAML exhibit significant advantages over those using VKF. VKF-based channel prediction is less effective since the sampled covariance matrix it relies on becomes inaccurate when only a small number of adaptation samples are used. Additionally, the hierarchical structure of MAML-based channel prediction results in a superior NMSE compared to the performance of both TL and MLP-based channel predictions.

\section{Discussion}
This section discusses key findings in channel prediction, including the training process and dataset consideration, as well as the impact of source tasks and pre-trained models.
\subsection{Training process and dataset consideration}
The training process and dataset requirements differ significantly between temporal channel prediction and environmental adaptation models due to their distinct objectives. Temporal channel prediction focuses on modeling time-dependent variations in channel conditions, which requires sequential data with strong temporal correlations. Models in this category, such as AR model and NNs, benefit from datasets that capture stable temporal dynamics. Sequence length selection is crucial in training these models: short sequences fail to capture long-term dependencies, while excessively long sequences increase computational complexity and may lead to gradient instability. High-mobility scenarios require adaptive sequence length selection to maintain performance. Moreover, datasets should include diverse mobility conditions with varying sequence lengths to ensure robust generalization. For RNNs, gradient clipping and layer-wise adaptive learning rates enhance training stability and mitigate the risk of overfitting, particularly when the dataset is limited.

In contrast, environmental adaptation models require datasets that reflect a wide range of environmental conditions to ensure robustness. Dataset diversity, covering various propagation environments (e.g., urban, suburban, indoor), significantly improves generalization. For example, models trained on datasets incorporating measurements from multiple frequency bands exhibit improved generalization, as they learn to adapt to diverse propagation characteristics compared to models trained on single-frequency band datasets. Additionally, data augmentation techniques improve model robustness; however, their effectiveness is dependent on the underlying channel characteristics and may lead to overfitting if not carefully implemented.

\subsection{Impact of source tasks and pre-trained models}
In the environmental adaptation, the selection of source tasks and pre-trained models plays a pivotal role in determining the effectiveness of both transfer learning and meta-learning models. For transfer learning, our experiments revealed that models pre-trained on datasets with environmental characteristics closely aligned with the target domain exhibit significantly better performance. Specifically, models pre-trained on urban macro (UMa) datasets, which cover large-scale urban environments with high-rise buildings and wide streets, demonstrated better generalization when adapted to new scenarios such as high-mobility vehicular networks in urban highways. In contrast, models pre-trained on UMi datasets, characterized by denser deployments and smaller cell coverage, showed limited adaptability to environments with large coverage areas, resulting in slower convergence and degraded performance. This highlights the critical role of environmental similarity between the source and target domains for transfer learning families, with the pre-training on dynamic environments like UMa enhancing model robustness.

For meta-learning, the diversity and structure of source tasks significantly influence the model’s adaptation capability. We observed that training with a heterogeneous set of source tasks led to faster adaptation and superior performance in previously unseen environments. Models trained with these diverse tasks adapted to new environments more efficiently compared to models exposed to homogeneous tasks. Interestingly, tasks involving high mobility contributed the most to adaptation efficiency, as they encouraged the model to develop robust, environment-agnostic feature representations. These findings show the importance of task diversity in source task selection for meta-learning families, guiding the development of more adaptable models for the real-world scenario.

\section{Challenges and Future Research Directions}
In this section, we discuss the possible challenges and the future research directions. We describe three challenges: multi-user MIMO-orthogonal frequency division multiplexing (OFDM) channel prediction, real-time channel prediction, and channel prediction with generative models. Then, we propose future research directions corresponding to each challenge.

\subsection{Multi-user MIMO-OFDM channel prediction}
Multi-user MIMO combined with OFDM is a key technology for modern wireless communication systems. However, channel prediction in multi-user MIMO-OFDM systems is challenging due to various reasons. In multi-user MIMO systems, there are multiple antennas at both the transmitter and receiver, and multiple UEs share the same resources. This leads to a high-dimensional and complex channel matrix that needs to be estimated and predicted. 
Managing and predicting inter-user interference is also challenging. Furthermore, since OFDM divides the signal into multiple orthogonal subcarriers, each subcarrier experiences different channel conditions. Predicting this accurately for multiple UEs and antennas adds another layer of complexity.

As our previous study in \cite{Ko2022} showed, for the single-user case, the subcarrier channels in OFDM can provide a large amount of training samples in a short period of time. 
We expect that meta-learning can be incorporated into the proposed technique in \cite{Ko2022} to address the multi-user MIMO-OFDM channel prediction problem, which is a future research direction.

\subsection{Real-time channel prediction}
Real-time channel prediction poses several challenges due to the need for immediate and accurate prediction within constrained time slots. The prediction model needs to be light-weight and computationally efficient for rapid responses. 
However, obtaining a sufficient amount of labeled data for real-time channel prediction remains challenging for advanced ML-based prediction approaches due to dynamic changes in environmental conditions, mobility, and interference in practice. Therefore, real-time channel prediction should be designed to adapt swiftly to these dynamic conditions.

To handle these challenges, we can consider light-weight NNs for the channel prediction \cite{Yang21_light}. Simple architecture, e.g., shallow feed-forward NNs or small-scale CNNs, can be effective for capturing relevant patterns in the channel data while keeping the model light-weight. In addition,  pruning and model quantization can be used to reduce the model size and storage requirements in the NN model. 
It would be an interesting research topic to investigate how to incorporate these approaches into advanced ML-based predictions for real-time channel prediction.

\subsection{Channel prediction with generative models}
For the environmental adaptation, generative models can be leveraged to predict the channel. As discussed in Section IV-C, GAN-based or VAE-based methods hold promise for enhancing channel prediction performance through data augmentation. 
However, GANs are susceptible to instability during the training phase and may exhibit limited diversity in generation due to their adversarial training architecture. Similarly, VAEs depend on a surrogate loss function, e.g., evidence lower bound (ELBO), where the ELBO may not directly correspond to minimizing the true data distribution's Kullback-Leibler (KL) divergence. To overcome these challenges, a diffusion model (DM) would be a promising approach for the channel prediction.

In contrast to GANs or VAEs, DMs adhere to a fixed learning procedure and incorporate latent variables with high dimensionality, which can generate high-fidelity samples. They utilize a Markov chain of diffusion steps to gradually introduce random noise to the data and then learn to reverse this diffusion process to generate desired data samples. 
Furthermore, the latent diffusion model (LDM), representing the state-of-art DM approach, would be used for the channel prediction using the cross-attention layers into its architecture.

\section{Conclusion}
In this paper, we first clarified two types of channel prediction approaches including temporal channel prediction and environmental adaptation and then described the model-based and ML-based channel predictions separately. The advanced ML-based channel prediction approaches, e.g., transfer learning, meta-learning, and attention mechanism, were elaborated to solve the limitations of training data, unseen tasks, and vanishing problems. Numerical results demonstrated that ML-based approaches outperform model-based methods, especially in environmental adaptation. We examined the training process, dataset properties, and the influence of source tasks and pre-trained models on channel prediction approaches. In addition, we discussed the future research directions to address the channel prediction challenges, which include the multi-user MIMO-OFDM channel prediction, real-time prediction, and channel prediction with generative models. Based on the future research directions in this paper, we believe that the advanced ML-based channel prediction approaches can also be applied to RIS-assisted MIMO systems, cell-free massive MIMO, or satellite communications for improving performance.

\section*{Acknowledgments}
This research was partly supported by Institute of Information \& communications Technology Planning \& Evaluation (IITP) under Open RAN Education and Training Program (IITP-2025-RS-2024-00429088) grant funded by the Korea government (MSIT) and by the National Science Foundation (NSF) under grants CNS-2212565 and CNS-2225577.

\bibliographystyle{IEEEtran}
\bibliography{IEEEcommag}

\section*{Biographies}

\begin{IEEEbiographynophoto}{Hwanjin Kim}
	(hwanjin@knu.ac.kr) received his Ph.D. degree in electrical engineering from KAIST in 2022. Since 2024, he is now the School of Electronics Engineering at Kyungpook National University as an assistant professor.
\end{IEEEbiographynophoto}

\begin{IEEEbiographynophoto}{Junil Choi}
	(junil@kaist.ac.kr) received his Ph.D. degree in electrical and computer engineering from Purdue University in 2015. He is now with the School of Electrical Engineering at KAIST as an associate professor.
\end{IEEEbiographynophoto}

\begin{IEEEbiographynophoto}{David J. Love}
	(djlove@purdue.edu) received his Ph.D. degree in electrical engineering from The University of Texas at Austin in 2004. He is now the Nick Trbovich Professor of the Elmore Family School of Electrical and Computer Engineering at Purdue University.
\end{IEEEbiographynophoto}

\end{document}